\documentclass[epj]{svjour}
\usepackage{graphicx,color}
\newcommand{\al}{$\alpha$}

\newcommand{\raa}{($\alpha$,$\alpha$)}

\newcommand{\crvi}{$^{46}$Cr}
\newcommand{\criv}{$^{54}$Cr}
\newcommand{\ruviii}{$^{88}$Ru}
\newcommand{\runull}{$^{90}$Ru}
\newcommand{\ruvi}{$^{96}$Ru}
\newcommand{\moiv}{$^{84}$Mo}
\newcommand{\movi}{$^{86}$Mo}
\newcommand{\moii}{$^{92}$Mo}
\newcommand{\zrnull}{$^{80}$Zr}
\newcommand{\zrii}{$^{82}$Zr}
\newcommand{\pdnull}{$^{90}$Pd}
\newcommand{\redreview}{}

\begin{document}
\title{
Yrast band in the heavy $N = Z$ nucleus $^{88}$Ru: $\alpha$-cluster approach
}
%
%
\author{Peter Mohr\inst{1,2}
\thanks{\emph{Email: mohr@atomki.mta.hu}} 
}                     
%
%
\institute{
  Institute for Nuclear Research (Atomki), H-4001 Debrecen, Hungary
  \and 
  Diakonie-Klinikum, D-74523 Schw\"abisch Hall, Germany
}
\date{Received: date / Revised version: date}
%
\abstract{
The yrast band in the heavy $N = Z$ nucleus $^{88}$Ru is studied in the
framework of the $\alpha$-cluster model in combination with double-folding
potentials. It is found that the excitation energies of the yrast band in
$^{88}$Ru can be nicely described within the $\alpha$-cluster approach using a
smooth and mildly $L$-dependent adjustment of the potential strength. This
result is similar to well-established $\alpha$-cluster states in nuclei with a
(magic core $\otimes$ $\alpha$) structure. Contrary, the yrast bands in
neighboring $N \ne Z$ nuclei deviate from such a typical $\alpha$-cluster
behavior. Finally, the $\alpha$-cluster model predicts reduced transition
strengths of about 10 Weisskopf units for intraband transitions between
low-lying states in the yrast band of $^{88}$Ru.
} 
\maketitle
\section{Introduction}
\label{sec:intro}

The structure of the heaviest accessible $N = Z$ nuclei has been subject of
interest over the last years, with one focus on the role of proton-neutron
pairing \cite{Frau14,Kim18}. Very recently, a detailed experimental
investigation of \ruviii\ became available which provided the excitation
energies of the yrast band up to $J^\pi = (14^+)$ and $E^\ast = 6949$ keV
\cite{Ceder20}. It was concluded that the observed yrast band in
\ruviii\ exhibits a band crossing ``that is significantly delayed compared
with the expected bahavior of a rotating deformed nucleus in the presence of a
normal isovector ($T = 1$) pairing field ... in agreement with theoretical
predictions for the presence of isoscalar neutron-proton pairing''. But such
an interpretation may also be affected by shape coexistence effects, and
despite several detailed studies \cite{Ceder20,Liu11,Hase04,Mar02,Fischer01} a
firm conclusion on the experimental verification of proton-neutron pairing
could not yet be reached. The relevance of shape coexistence
has also been pointed out in large-scale shell model calculations of heavy $N
= Z$ nuclei \cite{Zuker15}.

In the present work I use a completely different approach for the description
of the yrast band in \ruviii\ which is based on the simple two-body \al
-cluster model. In general, \al -clustering is a very well-known phenomenon in
nuclear physics \cite{Ren18,Hor12}. A lot of work has been done in the past for
doubly-magic cores, i.e.\ $^{212}$Po = $^{208}$Pb $\otimes$ \al , $^{44}$Ti =
$^{40}$Ca $\otimes$ \al , $^{20}$Ne = $^{16}$O $\otimes$ \al , and $^{8}$Be =
$^{4}$He $\otimes$ \al\ (see e.g.\ the recent review \cite{Ren18}), and
$^{94}$Mo = $^{90}$Zr $\otimes$ \al\ was often used to fill the wide gap of
missing stable doubly-magic core nuclei between $A \approx 40$ and $A > 200$
\cite{Ohk95}. In this gap, the \al -decay in the $N = Z$ system $^{104}$Te =
$^{100}$Sn $\otimes$ \al\ was also under study very recently
\cite{Cla20,Yang20,Aura18,Mohr07,Xu06}, and $^{52}$Ti = $^{48}$Ca $\otimes$
\al\ was investigated in \cite{Ohk20}.  A detailed introduction into the
nuclear cluster model is provided in a dedicated special issue of
Prog.\ Theor.\ Phys.\ Suppl.\ {\bf 132},
\cite{Ohk98,Mic98,Yam98,Sak98,Ueg98,Has98,Koh98,Toh98}.

In the last years, several successful attempts have been made to extend the
\al -cluster model towards semi-magic core nuclei (e.g., above $N = 50$
\cite{Mohr08}), and \al -cluster states have also been identified in non-magic
$^{46,54}$Cr \cite{Sou17,Mohr17}. Thus, it makes sense to apply the \al
-cluster model to \ruviii\ to see whether the properties of \ruviii\ can also
be described without explicit inclusion of proton-neutron pairing. It has to
be noted that the \al -cluster model and proton-neutron pairing are not
completely different approaches. It was already supposed in \cite{Hase04}
that the collaboration of the usual proton-proton and neutron-neutron pairing
with enhanced proton-neutron pairing in \ruviii\ may lead to \al -like
correlations, but unfortunately this idea remained qualitative in
\cite{Hase04} and was not detailed further.

The present study is organized as follows. Sec.~\ref{sec:fold} provides a
brief introduction into the chosen model. Sec.~\ref{sec:res} gives results for
\ruviii\ and some neighboring $N = Z$ and $N \ne Z$ nuclei. The results are
summarized in Sec.~\ref{sec:conc}. Experimental data are taken from the online
database ENSDF \cite{ENSDF}, except stated explicitly in the text.

\section{$\alpha$ clustering and folding potentials}
\label{sec:fold}
Details of the \al -cluster model are described in literature (e.g.,
\cite{Ohk98,Mic98,Yam98,Sak98,Ueg98,Has98,Koh98,Toh98}). Exactly the same
approach was used in my previous paper on $^{46,54}$Cr \cite{Mohr17}. Here I
briefly repeat important features of the model and its essential ingredients.

\subsection{Folding potential}
\label{sec:pot}
The interaction between the \al\ particle and the core nucleus is calculated
from the folding procedure with the widely used energy- and density-dependent
DDM3Y interaction $v_{\rm{eff}}$:
\begin{equation}
V_F(r) = 
	\int \int \, \rho_P(r_P) \,\rho_T(r_T) \,
        v_{\rm{eff}}(s,\rho,E_{\rm{NN}}) \; d^3r_P \; d^3r_T
\label{eq:fold}
\end{equation}
For details of the folding approach and the chosen interaction $v_{\rm{eff}}$,
see e.g.\ \cite{Sat79,Kob84,Mohr13}. The density $\rho_P$ of the \al -particle
is taken from the experimental charge density distribution
\cite{Vri87}. Obviously, for \ruviii\ = \moiv\ $\otimes$ \al\ the required
density of the unstable $N = Z$ nucleus \moiv\ is not available from
experiment. It has been seen recently, that the usage of microscopic
theoretical densities leads to very similar folding potentials
\cite{Mohr20}. In particular, Hartree-Fock-Bogolyubov densities using Skyrme
forces were used in the present approach {\redreview for all target densities
$\rho_T$}. These densities have been calculated by S.\ Goriely
and are provided as a part of the statistical model code TALYS \cite{TALYS}.
{\redreview The folding integral in Eq.~(\ref{eq:fold}) was solved in the
  so-called ``frozen-density'' approximation.}

The total interaction potential $V(r)$ is given by 
\begin{equation}
V(r) = V_N(r) + V_C(r) = \lambda \, V_F(r) + V_C(r)
\label{eq:vtot}
\end{equation}
where the nuclear potential $V_N$ is the double-folding potential $V_F$ of
Eq.~(\ref{eq:fold}) multiplied by a strength parameter $\lambda \approx 1.1 -
1.3$ \cite{Atz96,Mohr13}. $V_C$ is the Coulomb potential in the usual form of
a homogeneously charged sphere with the Coulomb radius $R_C$ chosen the same
as the root-mean-square radius $r_{\rm{rms}}$ of the folding potential $V_F$
which approximately follows $r_{\rm{rms}} \approx R_0 \times A_T^{1/3}$ with
$R_0 \approx 1.11 - 1.14$ fm for the nuclei under study. $A_T$ ($Z_T$, $N_T$)
will be used as the mass (charge, neutron) numbers of the ``target'' or core
nucleus in the \al -cluster model.

The strength parameter $\lambda$ is adjusted to reproduce the energies of the
bound states with $E < 0$ and quasi-bound states with $E > 0$ where $E = 0$
corresponds to the threshold of \al\ emission in the compound nucleus. In
general, $E = S_\alpha + E^\ast$ with the binding energy $S_\alpha$ of the
\al\ particle in the compound nucleus ($S_\alpha < 0$ for the nuclei under
study) and the excitation energy $E^\ast$. The number of nodes $N$ of the
bound state wave function $u_{NL}(r)$ was derived from the Wildermuth
condition
\begin{equation}
Q = 2N + L = \sum_{i=1}^4 (2n_i + l_i) = \sum_{i=1}^4 q_i
\label{eq:wild}
\end{equation}
where $Q$ is the number of oscillator quanta, $N$ is the number of nodes, and
$L$ is the relative angular momentum of the $\alpha$-core wave function. $q_i
= 2n_i + l_i$ are the corresponding quantum numbers of the nucleons in the
$\alpha$ cluster. Low-lying single-particle states around \ruviii\ are
dominated by the $g_{9/2}$ subshell, and thus I use $q_i = 4$, resulting in $Q
= 16$. This leads to an yrast band with nine positive-parity states with
$J^\pi$ from $0^+$ to $16^+$ for the even-even nuclei under study.

Further details on the formalism for the calculations of the folding
potentials $V_F(r)$, the resulting wave functions $u_{NL}(r)$, and reduced
transition strengths $B(E{\cal{L}})$ have been provided in earlier work
\cite{Mohr07,Mohr08,Atz96,Abe93,Hoy94}.

\subsection{{\redreview{Evidence for $\alpha$ clustering}}}
\label{sec:evi}
{\redreview{
    In light nuclei (e.g., $^{8}$Be = $^{4}$He $\otimes$ \al\ or $^{20}$Ne =
    $^{16}$O $\otimes$ \al ) \al -cluster states can be identified
    experimentally either by enhanced cross sections in \al -transfer
    experiments like ($^6$Li,$d$) or by large reduced widths $\theta_\alpha^2$
    in \al -induced reactions. In heavy nuclei the preformation factor $P$ in
    \al -decay can provide valuable information on \al -clustering of the
    decaying parent nuclei. However, for nuclei in the $A \approx 90$ mass
    region, the applicability of the above methods is limited. The lightest
    \al\ emitters are found slightly above the doubly-magic core $^{100}$Sn
    \cite{Aura18} which obviously excludes the $A \approx 90$ nuclei from \al
    -decay studies. Compared to light nuclei, \al -transfer studies like
    ($^6$Li,$d$) and the determination of reduced widths $\theta_\alpha^2$ are
    complicated for $A \approx 90$ nuclei by small cross sections because of
    the higher Coulomb barrier.

    The well-established \al -cluster states in light nuclei can be described
    as eigenstates in reasonably chosen \al -nucleus potentials (like
    e.g.\ the double-folding potential). The resulting wave functions of these
    eigenstates can be used to calculate $E2$ transition strengths within
    \al -cluster bands. In practice, these potentials show either a mild
    dependence on the angular momentum $L$ (e.g., \cite{Abe93}), or very
    special radial shapes of the potential like the $\cosh{}$ potential
    \cite{Buck93} or combinations of Woods-Saxon plus cubed Woods-Saxon
    potentials with further modifications \cite{Buck95,Sou15,Sou17} have to be
    chosen. Note that a mild angular momentum dependence is expected for local
    effective potentials like the folding potential of the present study
    \cite{Hor84,Mic95}.

    For nuclei in the $A \approx 90$ mass region, \al -clustering is usually
    assumed as soon as the excitation energies $E^\ast$ and $E2$ transition
    strengths for a rotational band can be described in the same way as
    explained for light nuclei in the previous paragraph
    \cite{Ohk95,Buck95}. Contrary, if the excitation energies of a band
    cannot be reproduced from reasonable \al -nucleus potentials, theses
    states are not considered as \al -cluster states.
    
    For double-folding potentials,
}}
typically a smooth decrease of the potential strength parameter $\lambda$ is
found with increasing excitation energy or increasing angular momentum
\cite{Abe93,Mohr08}. For intermediate mass nuclei around $N = 50$ an almost
linear decrease of $\lambda$ is found for the whole ground state band
\cite{Mohr08} whereas for the lightest nuclei the decreasing trend of
$\lambda$ may change to an increasing $\lambda$ for states above $L \approx 6$
\cite{Abe93}.

\subsection{{\redreview{$\alpha$-cluster properties of $^{96}$Ru}}}
\label{sec:ru96ref}
{\redreview{
    Most previous studies (e.g., \cite{Ohk95,Buck95}) in the $A \approx 90$
    mass region have focused on \al -cluster states in $^{94}$Mo = $^{90}$Zr
    $\otimes$ \al\ with the semi-magic $N_T = 50$ core
    $^{90}$Zr. Experimentally, the $0^+$ ground state and $2^+$ first excited
    state in $^{94}$Mo were populated in the $^{90}$Zr($^6$Li,$d$)$^{94}$Mo
    reaction \cite{Ful77}, thus confirming the \al -cluster properties of
    $^{94}$Mo. Based on a theoretical study, it has been pointed out in
    \cite{Sou15} that several nuclei with a ($N_T = 50$) $\otimes$
    \al\ structure show pronounced and very similar \al -clustering
    properties. Besides $^{94}$Mo, \al -clustering is suggested for $^{90}$Sr,
    $^{92}$Zr, $^{96}$Ru, and $^{98}$Pd in \cite{Sou15}. In particular, this
    finding has been confirmed recently for \ruvi\ by an \al -transfer
    experiment using the $^{92}$Mo($^{32}$S,$^{28}$Si)$^{96}$Ru reaction where
    the ground state band and few states of a low-lying negative-parity band
    in \ruvi\ were preferentially populated \cite{Datta18}.

    Based on the above theoretical and experimental arguments, the nucleus
    \ruvi\ will be used as a reference for \al -clustering in the
    following. \ruvi\ is preferred over $^{94}$Mo because the yrast band is
    completely identified up to $J^\pi = 16^+$ for \ruvi\ whereas for
    $^{94}$Mo levels are known only up to $J^\pi = (12^+)$. Note also that
    previous studies of $^{94}$Mo were restricted up to $J^\pi = 10^+$
    \cite{Buck95} and $J^\pi = 8^+$ \cite{Ohk95}, and both studies
    \cite{Ohk95,Buck95} disagreed on the location of the $J^\pi = 8^+$ state
    in $^{94}$Mo.
}}

\section{Results and Discussion}
\label{sec:res}
\subsection{$N = Z$ nucleus $^{88}$Ru = $^{84}$Mo $\otimes$ $\alpha$ and
  some general remarks}
\label{sec:ru88}

\begin{figure}[htb]
  \includegraphics[width=\columnwidth]{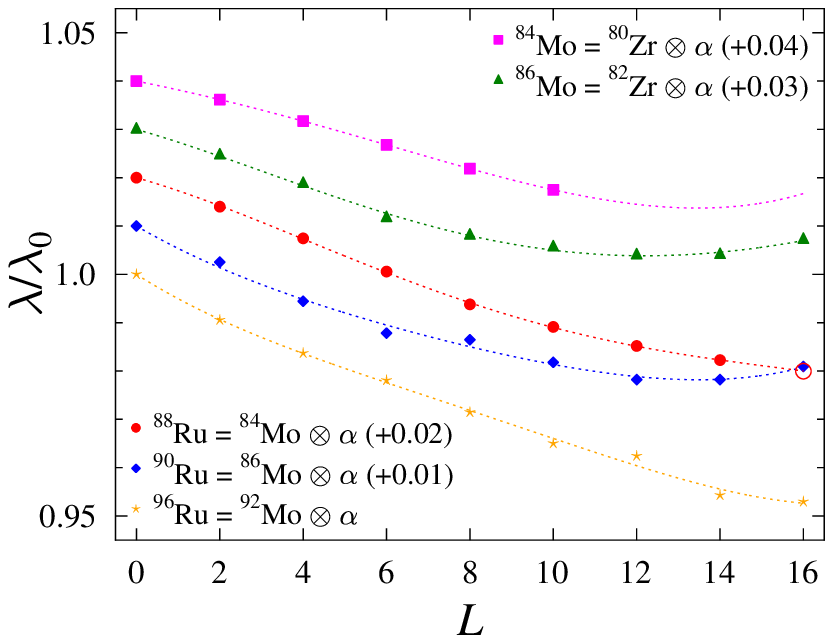}
\caption{
\label{fig:lambda}
Potential strength parameter $\lambda$ (normalized to $\lambda_0$ for the $L =
0$ ground states) for \ruviii\ = \moiv\ $\otimes$ \al\ in comparison to the
neighboring nuclei
\moiv\ = \zrnull\ $\otimes$ \al ,
\movi\ = \zrii\ $\otimes$ \al ,
\runull\ = \movi\ $\otimes$ \al , and
\ruvi\ = \moii\ $\otimes$ \al .
{\redreview{
    The red open circle shows the prediction for
    the experimentally unknown $J^\pi = 16^+$ state in \ruviii\ from
    Eq.~(\ref{eq:lam}) (see also Table \ref{tab:ruviii}).}}
For better visibility, the results for the different nuclei are shifted by
vertical offsets (0.01 per nucleus). Further discussion see text.
}
\end{figure}

Results for the potential strength parameter $\lambda$ of \ruviii\ are shown
in Fig.~\ref{fig:lambda}; numerical values are summarized in Table
\ref{tab:ruviii}. The obtained potential strength parameters $\lambda$ show a
mild, very smooth, and almost linear decrease from about $\lambda_0 \approx
1.27$ for the $0^+$ ground state down to about 1.22 for the $14^+$ state at
$E^\ast = 6949$ keV.
\begin{table*}
\caption{
\label{tab:ruviii}
  \al -cluster properties of \ruviii . Experimental excitation energies are
  taken from \cite{Ceder20}. Further discusssion see text.
}
\begin{center}
\begin{tabular}{crrrrccrr}
\hline
$J^\pi$ 
& \multicolumn{1}{c}{$E^\ast$} 
& $N$ & $L$ 
& \multicolumn{1}{c}{$\lambda$}
& \multicolumn{1}{c}{$J_R$} 
& \multicolumn{2}{c}{$B(E2,L\rightarrow L-2)$}
& \multicolumn{1}{c}{$\Gamma_\gamma$} \\
& (keV)
& & 
& 
& \multicolumn{1}{c}{(MeV\,fm$^3$)} 
& \multicolumn{1}{c}{(e$^2$\,fm$^4$)}
& \multicolumn{1}{c}{(W.u.)}
& \multicolumn{1}{c}{($\mu$eV)} \\
%
\hline
%
$0^+$   &       $0$     &  8   &  0   & 1.2663    & 352.4 
& -- & -- & -- \\
$2^+$   &       $616$   &  7   &  2   & 1.2587    & 350.2
& 203.7 &  8.8 &  15 \\
$4^+$   &      $1416$   &  6   &  4   & 1.2504    & 347.9
& 281.2 & 12.1 &  76 \\
$6^+$   &      $2380$   &  5   &  6   & 1.2417    & 345.5
& 285.2 & 12.3 & 196 \\
$8^+$   &      $3480$   &  4   &  8   & 1.2331    & 343.1
& 260.3 & 11.2 & 346 \\
$10^+$  &      $4543$   &  3   & 10   & 1.2272    & 341.5
&       &      &       \\
$12^+$  &      $5696$   &  2   & 12   & 1.2222    & 340.1
&       &      &       \\
$14^+$  &      $6949$   &  1   & 14   & 1.2185    & 339.1
&       &      &       \\
$16^+$ $^{\rm{a}}$
        &      $8370$   &  0   & 16   & 1.2156    & 338.2
&       &      &       \\
\hline
\multicolumn{9}{l}{\footnotesize{$^{\rm{a}}$Results for $16^+$ derived
    from the extrapolated $\lambda = 1.2156$ (see also
    Fig.~\ref{fig:lambda}).}} \\ 
\end{tabular}
\end{center}
\end{table*}

For a perfect reproduction of $\lambda(L)$, a 4$^{th}$-order fit has been made
(shown as dotted lines in Fig.~\ref{fig:lambda}):
\begin{equation}
  \lambda(L) = \lambda_0 \, \times \,
  \Bigl[ 1 + \sum_{i=1}^4 c_i L^i \Bigr]
\label{eq:lam}
\end{equation}
Eq.~(\ref{eq:lam}) provides an excellent description of $\lambda(L)$ for all
nuclei under investigation in the present work and has thus been used in all
cases.

Using Eq.~(\ref{eq:lam}), it is possible to extrapolate the experimentally
based potential strength parameters $\lambda$ of the $0^+$ to $14^+$ states to
the unobserved $J^\pi = 16^+$ state, leading to $\lambda(16^+) = 1.2156$ which
corresponds to an excitation energy of $E^\ast(16^+) \approx 8370$ keV.

It is interesting to note that the resulting volume integral $J_R$ of the
potential remains close to the recent ATOMKI-V1 global \al -nucleus potential
which was derived from elastic \raa\ scattering for heavy target nuclei at low
energies around the Coulomb barrier \cite{Mohr13}. This potential has also
turned out to be very successful in the prediction of \al -induced reaction
cross sections at astrophysically relevant sub-Coulomb energies \cite{Mohr20}.

Finally, an intrinsic uncertainty for the derived potential strength
parameters $\lambda$ has to be briefly mentioned; fortunately, the relevance
will turn out to be minor. The parameter $\lambda_0$ for the $0^+$ ground
states depends on the binding energy of the \al -particle in the compound
nucleus. Using the latest mass tables \cite{AME16-1,AME16-2}, the
uncertainties of the \al\ binding energies may reach up to almost 500 keV for
the unstable $N = Z$ nuclei under study. Obviously, a larger binding energy is
related to a larger $\lambda$ value. However, a shift in energy by 500 keV
corresponds only to a marginal change in $\lambda$ of about $7 \times
10^{-3}$. This minor uncertainty affects all $\lambda(L)$ in the same
direction and practically cancels out in the presentation and interpretation
of $\lambda/\lambda_0$ in Fig.~\ref{fig:lambda}. For completeness, the \al
-binding energies $S_\alpha$ and the corresponding $\lambda_0$ are listed in
Table \ref{tab:lambda} for all nuclei under investigation in the present
study.
\begin{table*}
\caption{
\label{tab:lambda}
  \al -binding energies $S_\alpha$ from \cite{AME16-1,AME16-2} and the
  resulting potential strength parameters $\lambda_0$ and volume integrals
  $J_R$ of the $0^+$ ground states.
}
\begin{center}
\begin{tabular}{cccccccrc}
\hline
\multicolumn{1}{c}{nucleus} 
& $Z$ & $N$ 
& \multicolumn{1}{c}{core} 
& $Z_T$ & $N_T$ 
& \multicolumn{1}{c}{$S_\alpha$ (MeV)}
& \multicolumn{1}{c}{$\lambda_0$}
& \multicolumn{1}{c}{$J_R$ (MeV\,fm$^3$)} \\
%
\hline
%
\moiv   &  $42$ & $42$ & \zrnull & $40$ & $40$ & $-2.235$ & 1.2800 & 357.6 \\
\movi   &  $42$ & $44$ & \zrii   & $40$ & $42$ & $-2.904$ & 1.2716 & 354.1 \\
\ruviii &  $44$ & $44$ & \moiv   & $42$ & $42$ & $-2.594$ & 1.2663 & 352.4 \\
\runull &  $44$ & $46$ & \movi   & $42$ & $44$ & $-3.198$ & 1.2583 & 348.9 \\
\ruvi   &  $44$ & $52$ & \moii   & $42$ & $50$ & $-1.697$ & 1.1942 & 327.7 \\
\hline
\end{tabular}
\end{center}
\end{table*}

In the following, the results for the $N = Z$ nucleus \ruviii\ will be
compared to \ruvi\ with its neutron-magic ($N_T = 50$) core
\moii\ (Sec.~\ref{sec:ru96}), to all neighboring $N \ne Z$ nuclei with $N \pm
2$ or $Z \pm 2$ (Secs.~\ref{sec:ru90}, \ref{sec:mo86}, and
\ref{sec:ru86pd90}), and to the lighter $N = Z$ nucleus
\moiv\ (Sec.~\ref{sec:mo84}).

\subsection{Nucleus $^{96}$Ru = $^{92}$Mo $\otimes$ $\alpha$ with
  neutron-magic core ($N_T = 50$)} 
\label{sec:ru96}
As a first comparison, we discuss the potential strength parameters
$\lambda(L)$ for \ruvi\ which is -- besides $^{94}$Mo \cite{Ohk95,Buck95} -- a
well-established \al -cluster nucleus with a $N_T = 50$ neutron-magic core in
this mass region. \ruvi\ is preferred here because the yrast band is
well-defined up to at least $J^\pi = 16^+$ whereas no high-spin members of the
yrast band are available for $^{94}$Mo
{\redreview{(see also Sec.~\ref{sec:ru96ref})}}.

The calculations for \ruvi\ in \cite{Mohr08} have been repeated to ensure that
the results are not affected by the present choice of the nucleon density of
\moii\ (see also Sec.~\ref{sec:pot}). In both cases, using either an average
density in \cite{Mohr08} or using the theoretical density from \cite{TALYS} in
the present study, it is found that the potential strength parameter $\lambda$
depends only mildly and almost linear on the angular momentum $L$. Such a
finding is typically interpreted as clear evidence for \al
-clustering. Interestingly, the $L$ dependence of $\lambda$ is very similar
for \ruvi\ and \ruviii , pointing also to a significant amount of \al
-clustering in the $N = Z$ nucleus \ruviii\ without magic core. Furthermore, a
practically identical behavior was found for $^{98}$Pd = $^{94}$Ru $\otimes$
\al , again with a neutron-magic core $N_T = 50$ \cite{Mohr08}.

\subsection{Neighboring $N = Z + 2$ nucleus $^{90}$Ru = $^{86}$Mo $\otimes$
  $\alpha$}
\label{sec:ru90}
Contrary to \ruviii\ and to the nuclei \ruvi\ and $^{98}$Pd with their
neutron-magic $N_T = 50$ cores which all show an almost linear decrease of the
potential strength parameter $\lambda(L)$, the $N \ne Z$ nucleus
\runull\ shows a clearly different behavior. The obtained values for $\lambda$
show a somewhat larger scatter, and the trend of decreasing $\lambda(L)$
changes to increasing $\lambda(L)$ for the largest $L = 14$ and
$16$. Interestingly, a similar trend for $\lambda(L)$ is also found for an
odd-parity band in \runull\ which can be interpreted with $Q = 17$ in
Eq.~(\ref{eq:wild}).

Both effects, a significantly larger scatter in $\lambda(L)$ and a
non-monotonic bahavior, are usually interpreted that the simple \al -cluster
model is less applicable for such cases. Such a finding is not very surprising
for \runull\ which is a nucleus without a neutron-magic core.

\subsection{Neighboring $N = Z + 2$ nucleus $^{86}$Mo = $^{82}$Zr $\otimes$
  $\alpha$}
\label{sec:mo86}
The results for the $N \ne Z$ nucleus \movi\ are practically identical to the
above results for \runull . In particular, there is also a clear trend of
increasing $\lambda(L)$ for the largest $L$ under study. The scatter of the
$\lambda(L)$ values is somewhat smaller than for \runull , but still above
the very smooth results for \ruviii\ and \ruvi .

Similar to \runull , also the $Q = 17$ odd-parity band in \movi\ shows a trend
of increasing $\lambda(L)$ for the largest $L$. However, because the $L = 15$
and $L = 17$ members are not known experimentally, the trend to increasing
$\lambda(L)$ for large $L$ is not as clear for \movi\ as for \runull .

\subsection{Neighboring $N = Z - 2$ nuclei $^{86}$Ru = $^{82}$Mo $\otimes$
  $\alpha$ and $^{90}$Pd = $^{86}$Ru $\otimes$ $\alpha$}
\label{sec:ru86pd90}
{\redreview{
    Unfortunately, the experimental information on these extremely
    neutron-deficient $N = Z - 2$ nuclei is very limited \cite{ENSDF} because
    experiments with these neutron-deficient species are very difficult. In
    particular, it was not possible up to now to determine properties of
    rotational bands in these nuclei. Thus, although highly desirable, it is
    not possible to analyze the $N = Z - 2$ nuclei \pdnull\ and $^{86}$Ru in
    the same way as their $N = Z + 2$ mirror nuclei
    \runull\ (Sect.~\ref{sec:ru90}) and \movi\ (Sect.~\ref{sec:mo86}).
}}

\subsection{Neighboring $N = Z$ nucleus $^{84}$Mo = $^{80}$Zr $\otimes$
  $\alpha$}
\label{sec:mo84}
At first view, the resulting potential strength parameters $\lambda(L)$ for
the $N = Z$ nucleus \moiv\ look very similar to the above results for \ruviii
. However, the observed yrast band in \moiv\ is known only up to $J^\pi =
(10^+)$ \cite{ENSDF,Mar02} which does not allow a strong statement on the
behavior of $\lambda(L)$ for large angular momenta $L$. The
$4^{th}$-order fit in Eq.~(\ref{eq:lam}) shows a trend to increasing
$\lambda(L)$ for large $L$, similar to the $N \ne Z$ nuclei
\movi\ and \runull , but different from the neighboring $N = Z$ nucleus
\ruviii . A complete determination of the yrast band in \moiv\ up to $J^\pi =
16^+$ would be very important to answer the open question whether
\moiv\ behaves really similar to its $N = Z$ neighbor \ruviii .

Unfortunately, experimental data are also very limited for the next $N = Z$
nucleus $^{92}$Pd, but very recently new data became available for $^{96}$Cd
\cite{Davies19}. Interestingly, the resulting $\lambda(L)$ for the $N = Z$
nucleus $^{96}$Cd also show a clear trend of increasing $\lambda(L)$ for
larger angular momenta $L$, i.e., similar to the $N \ne Z$ nuclei \movi\ and
\runull , but different to the $N = Z$ nucleus \ruviii .

\subsection{{\redreview{Comparison and interpretation of the results}}}
\label{sec:disc}
From the above results it is obvious that the yrast band in the $N = Z$
nucleus \ruviii\ can be nicely described within the \al -cluster
model. Similar to well-established \al -cluster nuclei like \ruvi , the
resulting potential strength parameters $\lambda(L)$ behave very regularly and
show a smooth and almost linear trend to decrease towards larger angular
momenta $L$. Contrary to the $N = Z$ nucleus \ruviii , the neighboring $N \ne
Z$ nuclei \movi\ and \runull\ show a different behavior with increasing
$\lambda(L)$ towards large $L$. Interestingly, such a behavior is also found
for the heavy $N = Z$ nucleus $^{96}$Cd, and no clear conclusion is possible
for the neighboring $N = Z$ nucleus \moiv . Also lighter $N \ne Z$ nuclei like
\crvi\ and \criv\ \cite{Mohr17} show the typical $N \ne Z$ behavior of
increasing $\lambda(L)$ for large $L$. Thus, the properties of the yrast band
in \ruviii\ seem to be somewhat extraordinary, and the interpretation of these
data requires special care.

The different models for \ruviii , i.e., \al -clustering vs.\ proton-neutron
pairing, can be further tested by comparing their predictions for reduced
$E2$ transition strengths. The \al -cluster model predicts $B(E2)$ of the
order of 10 Weisskopf units for intraband transitions between low-lying
members of the yrast band (see Table \ref{tab:ruviii}), and typically these
predictions are in rough agreement with experiment even without the usage of
effective charges and do not deviate by more than a factor of two from
experimental values \cite{Mohr08}. Contrary to the \al -cluster model, various
shell model approaches provide higher $E2$ strengths of 20 or even more
Weisskopf units \cite{Hase04,Zuker15,Kan17}. Very high values, even exceeding
50 Weisskopf units, were found from the consideration of proton-neutron
pairing for the $B(E2,8^+\rightarrow 6^+)$, $B(E2,6^+\rightarrow 4^+)$, and
$B(E2,4^+\rightarrow 2^+)$ transitions in the shell model \cite{Kan17} whereas
the \al -cluster model predicts almost constant $B(E2)$ for all
transitions. Thus, a measurement of transition strengths could provide further
insight in the applicability of the different models for \ruviii . This holds
in particular for transition strengths between excited states which are
however extremely difficult to measure for \ruviii . All present predictions
of $B(E2)$ strengths from the shell model \cite{Hase04,Zuker15,Kan17} and from
the \al -cluster model in this study lead to lifetimes far below nanoseconds
in the yrast band and are thus compatible with the observation of
$\gamma$-$\gamma$ coincidences within the experimental nanosecond timing
window of HPGe detectors \cite{Ceder20}.

\section{Summary and Conclusions}
\label{sec:conc}
The \al -cluster model is able to reproduce the excitation energies of the
recently measured \cite{Ceder20} yrast band in the heavy $N = Z$ nucleus
\ruviii . This provides an alternative approach to previous studies which had
a focus on special proton-neutron pairing in $N = Z$ nuclei which was required
to describe the observed delayed band crossing, compared to neighboring $N \ne
Z$ nuclei.
{\redreview{
    Both approaches -- the \al -cluster model on the one side and the shell
    model in combination with proton-neutron pairing on the other side -- are
    able to reproduce the observed excitation energies of the yrast band in
    \ruviii .
}}

The experimental determination of radiation widths or absolute $E2$
transition strengths could be one way to test the reliability of the
predictions from the different models and the applicability to \ruviii
. Without a clear experimentally based preference for one of the models, any
conclusion on the experimental verification of proton-neutron pairing in
\ruviii\ or on significant \al -clustering in \ruviii\ has to remain
tentative.

\section*{Acknowledgments}
I thank Zs.\ F\"ul\"op, Gy.\ Gy\"urky, G.\ G.\ Kiss, T.\ Sz\"ucs, and
E.\ Somorjai for longstanding encouraging discussions on \al -nucleus
potentials.
This work was supported by NKFIH (NN128072, K120666), and by the
{\'U}NKP-19-4-DE-65 New National Excellence Program of the Ministry of Human
Capacities of Hungary.



\begin{thebibliography}{}
%
\bibitem{Frau14}
  S.\ Frauendorf and A.\ O.\ Macchiavelli,
  Prog.\ Part.\ Nucl.\ Phys.\ {\bf 78}, 24 (2014).
%
\bibitem{Kim18}
  Y.\ H.\ Kim, M.\ Rejmund, P.\ Van Isacker, and A.\ Lemasson,
  Phys.\ Rev.\ C {\bf 97}, 041302(R) (2018).
%
\bibitem{Ceder20}
  B.\ Cederwall {\it et al.},
  Phys.\ Rev.\ Lett.\ {\bf 124}, 062501 (2020).
%
\bibitem{Liu11}
  Liu XuDong, Shi Yue, Xu FuRong,
  Sci.\ China {\bf 54}, 1811 (2011).
%
\bibitem{Hase04}
  M.\ Hasegawa, K.\ Kaneko, T.\ Mizusaki, and S.\ Tazaki,
  Phys.\ Rev.\ C {\bf 69}, 034324 (2004).
%
\bibitem{Mar02}
  N.\ Marginean {\it et al.},
  Phys.\ Rev.\ C {\bf 65}, 051303(R) (2002).
%
\bibitem{Fischer01}
  S.\ M.\ Fischer {\it et al.},
  Phys.\ Rev.\ Lett.\ {\bf 87}, 132501 (2001).
%
\bibitem{Zuker15}
  A.\ P.\ Zuker, A.\ Poves, F.\ Nowacki, and S.\ M.\ Lenzi,
  Phys.\ Rev.\ C {\bf 92}, 024320 (2015).
%
\bibitem{Ren18}
  Zhongzhou Ren and Bo Zhou,
  Front.\ Phys.\ {\bf 13}, 132110 (2018).
%
\bibitem{Hor12}
H.\ Horiuchi, K. Ikeda, K.\ Kat{\={o}},
Prog.\ Theor.\ Phys.\ Suppl.\ {\bf 192}, 1 (2012).
%
\bibitem{Ohk95}
S.\ Ohkubo,
Phys.\ Rev.\ Lett.\ {\bf 74}, 2176 (1995).
%
\bibitem{Cla20}
  R.\ M.\ Clark, A.\ O.\ Macchiavelli, H.\ L.\ Crawford, P.\ Fallon,
  D.\ Rudolph , A.\ S{\aa}mark-Roth, M.\ Campbell, M.\ Cromaz, C. Morse, and
  C. Santamaria,
  Phys.\ Rev.\ C {\bf 101}, 034313 (2020).
%
\bibitem{Yang20}
  Shuo Yang, Chang Xu, Gerd R{\"o}pke, Peter Schuck, Zhongzhou Ren, Yasuro
  Funaki, Hisashi Horiuchi, Akihiro Tohsaki, Taiichi Yamada, and Bo Zhou,
  Phys.\ Rev.\ C {\bf 101}, 024316 (2020).
%
\bibitem{Aura18}
  K.\ Auranen {\it et al.},
  Phys.\ Rev.\ Lett.\ {\bf 121}, 182501 (2018).
%
\bibitem{Mohr07}
P.\ Mohr,
Europ.\ Phys.\ J.\ A {\bf 31}, 23 (2007).
%
\bibitem{Xu06}
Chang Xu and Zhongzhou Ren,
Phys.\ Rev.\ C {\bf 74}, 037302 (2006).
%
\bibitem{Ohk20}
  S.\ Ohkubo,
  Phys.\ Rev.\ C {\bf 101}, 041301(R) (2020).
%
\bibitem{Ohk98}
S.\ Ohkubo, M.\ Fujiwara, P.\  E.\ Hodgson,
Prog.\ Theor.\ Phys.\ Suppl.\ {\bf 132}, 1 (1998).
%
\bibitem{Mic98}
F.\ Michel, S.\ Ohkubo, G.\ Reidemeister,
Prog.\ Theor.\ Phys.\ Suppl.\ {\bf 132}, 7 (1998).
%
\bibitem{Yam98}
T.\ Yamaya, K.\ Katori, M.\ Fujiwara, S.\ Kato, S.\ Ohkubo,
Prog.\ Theor.\ Phys.\ Suppl.\ {\bf 132}, 73 (1998).
%
\bibitem{Sak98}
T.\ Sakuda and S.\ Ohkubo,
Prog.\ Theor.\ Phys.\ Suppl.\ {\bf 132}, 103 (1998).
%
\bibitem{Ueg98}
E.\ Uegaki,
Prog.\ Theor.\ Phys.\ Suppl.\ {\bf 132}, 135 (1998).
%
\bibitem{Has98}
M.\ Hasegawa,
Prog.\ Theor.\ Phys.\ Suppl.\ {\bf 132}, 177 (1998).
%
\bibitem{Koh98}
S.\ Koh,
Prog.\ Theor.\ Phys.\ Suppl.\ {\bf 132}, 197 (1998).
%
\bibitem{Toh98}
A.\ Tohsaki,
Prog.\ Theor.\ Phys.\ Suppl.\ {\bf 132}, 213 (1998).
%
\bibitem{Mohr08}
  P.\ Mohr,
  The Open Nuclear and Particle Physics Journal {\bf 1}, 1 (2008);
  {\it{arXiv:0803.2202}}.
%
\bibitem{Sou17}
M.\ A.\ Souza and H.\ Miyake,
Europ.\ Phys.\ J.\ A {\bf 53}:146 (2017).
%
\bibitem{Mohr17}
P.\ Mohr,
Europ.\ Phys.\ J.\ A {\bf 53}:209 (2017).
%
\bibitem{ENSDF}
Online database ENSDF,
available at {\it{www.nndc.bnl.gov/ensdf}};
based on \cite{Singh09,Negret15,McCut14,Browne97,Abri08}.
%
\bibitem{Singh09} 
  B.\ Singh,
  Nuclear Data Sheets {\bf 110}, 2815 (2009).
%
\bibitem{Negret15} 
  A.\ Negret and B.\ Singh,
  Nuclear Data Sheets {\bf 124}, 1 (2015).
%
\bibitem{McCut14} 
  E.\ A.\ McCutchan and A.\ A.\ Sonzogni,
  Nuclear Data Sheets {\bf 115}, 135 (2015).
%
\bibitem{Browne97} 
  E.\ Browne,
  Nuclear Data Sheets {\bf 82}, 379 (1997).
%
\bibitem{Abri08} 
  D.\ Abriola, A.\ A.\ Sonzogni,
  Nuclear Data Sheets {\bf 109}, 2501 (2008).
%
\bibitem{Sat79}
G.\ R.\ Satchler and W.\ G.\ Love,
Phys.\ Rep.\ {\bf 55}, 183 (1979).
%
\bibitem{Kob84}
A.\ M.\ Kobos, B.\ A.\ Brown, R.\ Lindsay, and G.\ R.\ Satchler,
Nucl.\ Phys.\ {\bf A425}, 205 (1984).
%
\bibitem{Mohr13}
P.\ Mohr, G.\ G.\ Kiss, Zs.\ F\"ul\"op, D.\ Galaviz, Gy.\ Gy\"urky, 
E.\ Somorjai,
At.\ Data Nucl.\ Data Tables {\bf 99}, 651 (2013).
%
\bibitem{Vri87}
  H.\ de Vries, C.\ W.\ de Jager, and C.\ de Vries,
  Atomic Data and Nuclear Data Tables {\bf 36}, 495 (1987).
%
\bibitem{Mohr20}
  P.\ Mohr, Zs.\ F\"ul\"op, Gy.\ Gy\"urky, G.\ G.\ Kiss, and T.\ Sz\"ucs,
  Phys.\ Rev.\ Lett., submitted.
%
\bibitem{TALYS}
  A.\ J.\ Koning, S.\ Hilaire, S.\ Goriely,
  computer code TALYS, version 1.8,
  {\it{http://www.talys.eu}};
  A.\ J.\ Koning, S.\ Hilaire, and M.\ C.\ Duijvestijn,
  AIP Conf.\ Proc.\ \textbf{769}, 1154 (2005).
%
\bibitem{Atz96} 
  U.\ Atzrott, P.\ Mohr, H.\ Abele, C.\ Hillenmayer, and
  G.\ Staudt,
  Phys.\ Rev.\ C {\bf 53}, 1336 (1996).
%
\bibitem{Abe93}
H.\ Abele and G.\ Staudt,
Phys.\ Rev.\ C {\bf 47}, 742 (1993).
%
\bibitem{Hoy94}
F.\ Hoyler, P.\ Mohr, G.\ Staudt,
Phys.\ Rev.\ C {\bf 50}, 2631 (1994).
%
\bibitem{Buck93}
  B.\ Buck, A.\ C.\ Merchant, S.\ M.\ Perez,
  At.\ Data Nucl.\ Data Tables {\bf 54}, 53 (1993).
%
\bibitem{Buck95}
  B.\ Buck, J.\ C.\ Johnston,  A.\ C.\ Merchant, S.\ M.\ Perez,
  Phys.\ Rev.\ C {\bf 51}, 559 (1995).
  %
\bibitem{Sou15}
  M.\ A.\ Souza and H.\ Miyake,
  Phys.\ Rev.\ C {\bf 91}, 034320 (2015).
%
\bibitem{Hor84}
  H.\ Horiuchi,
  Proc.~{\it{Int.\ Conf.\ on Clustering Aspects of Nuclear Structure and
      Nuclear Reactions}}, Chester, UK, July 23--27, 1984,
  Eds.\ J.\ S.\ Lilley and M.\ A.\ Nagarajan,
  Springer Book {\it{ https://www.springer.com/de/book/9789027720023}}.
%
\bibitem{Mic95}
  F.\ Michel, G.\ Reidemeister, and Y.\ Kond\=o,
  Phys.\ Rev.\ C {\bf 51}, 3290 (1995).
%
\bibitem{Ful77}
  H.\ W.\ Fulbright, C.\ L.\ Bennett, R.\ A.\ Lindgren, R.\ G.\ Markham,
  S.\ C.\ McGuire, G.\ C.\ Morrison, U.\ Strohbusch, J.\ T{\"o}ke,
  Nucl.\ Phys.\ {\bf A284}, 329 (1977).
%
\bibitem{Datta18}
  U.\ Datta {\it et al.},
  AIP Conf.\ Proc.\ {\bf 2038}, 020020 (2018).
%
\bibitem{AME16-1}
  W.\ J.\ Huang, G.\ Audi, M.\ Wang, F.\ G.\ Kondev, S.\ Naimi, and X.\ Xu,
  Chin.\ Phys.\ C {\bf 41}, 030002 (2017).
%
\bibitem{AME16-2}
  M.\ Wang, G.\ Audi, F.\ G.\ Kondev, W.\ J.\ Huang, S.\ Naimi, and X.\ Xu,
  Chin.\ Phys.\ C {\bf 41}, 030003 (2017).
%
\bibitem{Davies19}
  P.\ J.\ Davies {\it et al.},
  Phys.\ Rev.\ C {\bf 99}, 021302(R) (2019).
%
\bibitem{Kan17}
  K.\ Kaneko, Y.\ Sun, G.\ de Angelis,
  Nucl.\ Phys.\ {\bf A957}, 144 (2017).
%
\end{thebibliography}
\end{document}